\documentstyle[aps,12pt]{revtex}
\begin{document}
\draft
\preprint{ subm. to PRC, \today}
\normalsize
%\tightenlines
\title{Flow Study in Relativistic Nuclear
Collisions by Fourier Expansion of Azimuthal Particle Distributions}

\author{S.~Voloshin$^{1,}$\footnote{ On leave from Moscow
Engineering Physics Institute,
Moscow, 115409,  Russia}
, Y.~Zhang$^2$}
\address{$^1$ University of Pittsburgh, Pittsburgh, PA 15260}
\address{$^2$ State University of New York, Stony Brook, NY 11794}
\date{Received 2 June 1994}
\maketitle

\begin{abstract}
We propose a new method to study transverse flow effects in
relativistic nuclear collisions by Fourier analysis of the azimuthal
distribution on an event-by-event basis in relatively narrow rapidity
windows. The distributions of Fourier coefficients provide direct
information on the magnitude and type of flow. Directivity
 and two dimensional sphericity tensor, widely used to analyze
flow, emerge naturally in our approach, since they correspond to the
distributions of the first and second harmonic coefficients,
respectively. The role of finite particle fluctuations and particle
correlations is discussed.
\end{abstract}
\pacs{PACS number: 25.75.+r}

%\narrowtext

\section{Introduction} \label{sec1}

In the study of specific processes such as flow phenomena, it is very
important to find variables which are both easy to work with and have
clear physical interpretations. Many different methods were
proposed for the study of flow effects in relativistic nuclear
collisions, of which the most commonly used are directivity~\cite{ldo}
and sphericity (three dimensional~\cite{ldg} or two dimensional~\cite{lo1})
 tensor.
In the current paper we propose to use a more general method, which
includes as a part both directivity and two dimensional
sphericity methods in a natural way and gives a clear
physical meaning to the analysis.
Moreover, it provides ways to investigate
complicated event shapes, which may not be described by an ellipsoid
through three dimensional sphericity analysis.

There exist two different approaches in flow analysis.
The first one is to  fit  $p_t$ and/or $dN/dy$
distributions of different particles by assuming local thermal
equilibrium and hydro-dynamical expansion.
In this case some information about initial conditions,
equation-of-state, expansion dynamics,
freeze-out temperature can be extracted.
(For recent analysis of such type of
 CERN SPS and BNL AGS data see~\cite{sps,ags}).
In spite of sometimes very good description
of experimental data this approach
has essential plague: it is very model dependent. Just the very existing of
thermal equilibrium is still not well established.
The alternative approach is to study the azimuthal event shapes
(which are sensitive to flow)
from experimental data without any model in mind. All the methods
mentioned in the first paragraph including the method we are going to
discuss in this paper are falling into this category.
The information provided in this way can be used for comparison with
different models: hydro-dynamical, statistical, cascade,
multi-string model, etc.
The study of a possible azimuthal anisotropy of the multiparticle production
in high energy nucleus-nucleus collisions just has begun, and first
indications of it have been seen at AGS~\cite{l802},
where the anisotropy of particle production in one region was
noted to correlate with the reaction plane angle defined in another region.

In our approach we investigate the shapes of azimuthal distributions,
and try to detect and study of anisotropy of different kinds.
The origin of the anisotropy could be different:
hydro-dynamical flow due to pressure gradients,
shadowing, both, etc.
What they have in common is some collective behavior in the
evolution of multiparticle production process.
Consequently, we often use the term transverse collective flow
for the phenomena we study,
although we do not necessarily mean the hydro-dynamical flow.
Our method is most useful for detecting an {\em anisotropic}
collective flow. Though it can be used for a study of radial flow as well
(see the short discussion in the Section~\ref{sec2})
along with other methods
such as an analysis of $p_t$ spectra.
In order to detect complex three dimensional event shapes, we slice
longitudinal variables, e.g., (pseudo)-rapidity, into different windows,
and perform Fourier analysis in each window. Fourier coefficients of
different harmonics reflect the different type of anisotropy (transverse
collective flow).
A three dimensional event shape can be obtained by correlating
and combining the
Fourier coefficients in different longitudinal windows.
Some ideas of this approach have
been employed to analyze Au on Au data of E877 collaboration at BNL AGS,
when for the first time in this energy range transverse flow signals
have been observed~\cite{lhw,l877}.

We present the Fourier expansion method in Section~\ref{sec2}.
We study the fluctuations due to finite particle multiplicity in
section~\ref{sec3}. In section~\ref{sec4}, we address correlation
between different windows and different harmonics. Finally, relations
to other existing methods are discussed in section~\ref{sec5}.

\section{Fourier expansion of azimuthal distributions} \label{sec2}

The particle azimuthal distributions can be constructed from different
quantities such as transverse momentum, multiplicity, or transverse
energy in relatively narrow (pseudo)-rapidity windows. To get more
detail of the event shapes one would like to divide the whole
(pseudo)-rapidity acceptance into more windows. But the limitation is
that there should be sufficient number of particles in each window. We
will discuss this problem in next section.  If needed, it is possible to
introduce different weights for different particles, as is often done in
the analysis of flow phenomena at lower energies.  It is also possible
to study specific functions of $p_t$, multiplicity, etc; for example, if
one wants to enhance the contribution from particles with high $p_t$,
one can study $ p_t^2 (\phi)$ instead of $ p_t (\phi)$, or select
particles with $ p_t $ larger than certain value. In this way the method
can also be used for a study of radial flow. To study flow effects in
future RHIC experiments, where jet production can be important and will
be a background to flow study, one can
chose the range $p_t < c \langle p_t^{jet} \rangle$,
where constant $c$ can be chosen from experiments.
Or one can do just the opposite to study anisotropy due to
jets. With particle identification, one can get the
azimuthal function for certain type of particles, e.g., analyze protons
and pions separately.
In the following to be specific we generally discuss
an azimuthal distribution of transverse
momentum in certain rapidity window, but the procedure for other
variables remains the same.

We denote the azimuthal distribution function of the quantity
under study with $r(\phi)$, {\it i.e.}, $dp_T(\phi) / d\phi$, where
$p_T(\phi)$ is the total transverse momentum of particles emitted
at azimuthal angle $\phi$. The function $r(\phi)$ can be constructed from
experimental data event by event and be written in the form of Fourier
expansion since $r(\phi)$ is a periodical function:
\begin{equation}
r(\phi)=\frac{x_0}{2\pi} + \frac{1}{\pi}
\sum_{n=1}^{\infty}[x_n \cos (n\phi)+y_n \sin (n\phi)].
\end{equation}
The coefficients in the Fourier expansion of $r(\phi)$ are integrals of
$r(\phi)$ with weights proportional to $\cos (n\phi)$ or $\sin (n\phi)$.
For the case of a finite number of particles, the integrals become
simple sums over particles found in the appropriate rapidity window:
\begin{equation}
x_n = \int_0^{2\pi} r(\phi) \cos (n\phi) d \phi
    = \sum_{\nu} r_{\nu} \cos (n\phi_{\nu}),
\label{ecc}
\end{equation}
\begin{equation}
y_n = \int_0^{2\pi} r(\phi) \sin (n\phi) d \phi
    = \sum_{\nu} r_{\nu} \sin (n\phi_{\nu}),
\label{esc}
\end{equation}
where $\nu$ runs over all particles, and $\phi_{\nu}$ is the azimuthal
angle of the $\nu$-th particle. Without any flow and neglecting
fluctuations (we study the fluctuations in Section~\ref{sec3}) the
function $r(\phi)$ is constant, $ r(\phi) =
x_0/(2\pi) = 1/(2\pi)\,\sum_{\nu} r_{\nu}.$
All Fourier coefficients except $a_0$ are zero.

The appearance of transverse anisotropic flow in nuclear collision
is expected from a non-zero value of an impact parameter.
We define the longitudinal or beam direction as $z$-axis,
and the transverse plane as $x$-$y$ plane.
%In some of the discussion, the $x$-axis is in impact
%parameter direction. We will make a note when we use this definition,
%but in general $x$ and $y$ are treated the same.
The impact parameter vector in $x$-$y$ plane
(from the center of the target nucleus to the center of the beam nucleus)
together with $z$-axis defines the reaction
plane, and the reaction plane angle $\psi_r \; (0\leq \psi_r \leq 2\pi)$
is the angle between $x$-axis and reaction plane. Each non-zero pair of
the Fourier coefficients $x_n$ and $y_n$ give the value of non-zero
component of the corresponding harmonic, defined as
$v_n=\sqrt{x_n^2+y_n^2}$, and the angle $\psi_n$ ($0 \leq \psi_n < 2\pi
/n$) of the ``$n$-th type'' flow.
\begin{equation}
x_n=v_n \cos(n\psi_n),
\end{equation}
\begin{equation}
y_n=v_n \sin(n\psi_n).
\end{equation}

If flow exists, $r(\phi)$ is no longer a constant,
and the shape of the distribution
is no longer a circle centered at zero (see Fig.~\ref{ffl}).
Then the first harmonic coefficients correspond to an overall shift
of the distribution in the transverse plane;
such flow we call directed flow. The magnitude of flow
is $v_1$. With $r(\phi)$ being the transverse momentum distribution,
coefficients
$x_1$ and $y_1$ are $x$ and $y$ components of the total transverse
momentum of the particles produced in the window, while $v_1$ is the
magnitude of the total vector sum of transverse momenta.
In the case of negligible fluctuations the direction of flow due to symmetry
is to coincide with the reaction
plane angle $\psi_1 = \psi_r$ (repulsive flow),
or point in the opposite direction $\psi_1 = \psi_r + \pi$ (attractive flow).

The non-zero second harmonic describes the eccentricity of an
ellipse-like distribution.  If one approximates the distribution by an
ellipse as shown in Fig.~\ref{ffl}, then the second coefficient $v_2$
carries information on the magnitude of the eccentricity.
Quantitatively it is the difference between the major and the minor
axis. The orientation of major axis $\psi_2$ (or,  $\psi_2+\pi$, which
give the same orientation for an ellipse) can be only $\psi_r$ or
$\psi_r +\pi/2$. In the case $\psi_2=\psi_r$, the major axis lies
within the reaction plane; while $\psi_2=\psi_r+\pi/2$
indicates that the orientation of
the ellipse is perpendicular to the reaction plane, which is the case
for squeeze-out flow and may be expected in mid-rapidity window.

Coefficient $v_3$ could be non-zero for asymmetric nuclear collisions
and represent the asymmetry of the flow due to different sizes of
colliding nuclei, if the distribution can be approximately described as
triangle-type distribution.
In this case the symmetry dictates that the flow angle $\psi_3$
(or one of $\psi_3+2n\pi/3$, $n=1,2$) gets value $\psi_r$ or $\psi_r+\pi/3$.

Forth -- a rectangle-type deformation amplitude $v_4$ could be non-zero
for rapidity windows close to the center of mass, where both
squeeze-out, emitting preferentially more particles in the direction
perpendicular to the reaction plane, and ``side-splash'' effects can be
important. The flow angle $\psi_4$ would be $\psi_r$ if the origin of
the rectangle-type flow is the case as discussed above. But $\psi_4 =
\psi_r+\pi/4$ can not be excluded from the naive symmetric reaction
plane argument.

We see that Fourier coefficients corresponding to different harmonics
have a clear physical meaning. Perhaps the most interesting and unique
feature of this approach would be to correlate flow effects between
different types. But before trying to do that, we have to
answer the question of how to deal with the finite multiplicity
fluctuations, which have always been a big obstacle in flow analysis.
Because it is difficult to resolve the reaction plane angle event by
event due to fluctuations, all methods based on the determination of the
reaction plane on an event-by-event basis remains rather ambiguous. It
becomes very important to detect flow without using the information on
the exact position of the reaction plane in each event. When consider
the fluctuations, the value of $v_n$ becomes a distribution instead of a
single number for events of the same characterization.
Below we  show that
the distribution in $v_n$ is sensitive to anisotropic flow and can be
used for flow detection.
Information about the reaction plane position could be used if one
considers the relative angular distance between the reaction plane and
some other plane, also defined in the same event.  For analysis of this
kind we propose to study the correlations between reactions plane angles
defined in different windows, or between reaction plane angles derived,
possibly, from the same window but using different harmonics.

\section{Finite multiplicity fluctuations} \label{sec3}

In this section we discuss problems arising in the analysis due to
finite multiplicity fluctuations. Since two particle correlations are
weak~\cite{l814}, below we treat all particles in the same event as
being totally independent following certain distribution with variance
$\sigma_0^2$. If two particle correlations are non-negligible we expect
larger fluctuations, but the general procedure remains valid.  Although
we study finite multiplicity fluctuations, we consider the case when the
total number of particles in selected window is relatively large ($N\gg
1$).  Under these assumptions, in the absence of flow the probability
distribution of vectors $(v_1,\psi_1)$ is Gaussian from the central
limit theorem:
\begin{equation}
\frac{d^2w}{dv_1 d\psi_1}=\frac{1}{2 \pi \sigma^2}
\exp (-\frac{v_1^2}{ 2 \sigma^2}),
\label{ev1}
\end{equation}
with the peak of the distribution centered at the center of the
coordinates. The variance of the distribution is:
\begin{equation}
\sigma^2=\sigma^2_x=\sigma^2_y
=N\,\langle r_i^2\rangle \langle \cos^2(\phi_i) \rangle
=\frac{N\langle r_i^2\rangle}{2}
\equiv N\,\sigma_0^2,
\label{evar}
\end{equation}
where $N$ is the number of particles in the window under study.

For simplicity, we begin our study with flow of the first type, namely
directed (``side-splash'') flow.
With the appearance of flow, each event can be
characterized by the magnitude of flow and the flow angle.
For events with the
same parameters of flow the center of the distribution (\ref{ev1}) will
be shifted to the point characterized by $(\tilde{v}_1, \psi_r)$ in the
polar coordinate system.  It is reasonable to assume that in shifted
distributions the statistical fluctuations (\ref{evar}) are the same.
We have chosen here $\tilde{v}_1$ as the parameter
relevant to the magnitude of flow, which is directly related to the
impact parameter of the collision.  On the other hand, it is known from
Monte Carlo studies that the impact parameter is strongly correlated
with global variables used to measure the centrality of events
(transverse energy, multiplicity, zero degree energy, etc.).  Using
these correlations one can select events with the same magnitude of flow
experimentally. Below we assume that the distribution in $\tilde{v}_1$
is narrow and can be treated as a constant for events within the same
centrality bins.

The distribution in the flow angle $\psi_1$ should be flat. That can be
used experimentally to justify the equalization and correction applied
to detectors. We can integrate over $\psi_1'=\psi_1-\psi_r$, which gives:
\begin{equation}
\frac{dw}{v_1\,dv_1}=
\frac{1}{2\pi \sigma^2}\int_{0}^{2\pi}d\psi_1'
\exp (-\frac{\tilde{v}_1^2 + v_1^2
-2\tilde{v}_1 v_1\cos(\psi_1')}
{2\sigma^2})=\frac{1}{\sigma^2}\exp
(-\frac{\tilde{v}_1^2+v_1^2}{2\sigma^2})
I_0(\frac{\tilde{v}_1 v_1}{\sigma^2}),
\label{eb}
\end{equation}
where $I_0$ is the modified Bessel function.  The shape of the
distribution (\ref{eb}) for a few values of parameters are shown in
Fig.~\ref{feb}, where we plot $dw/xdx$ using the notation $x \equiv
v_1/\sigma$ and $\tilde{x} \equiv \tilde{v}_1/\sigma$. The extrema are
defined by the solution of the equation
\begin{equation}
\tilde{v}_1 \, I_1(v_1\tilde{v}_1/\sigma^2 )
-v_1\, I_0(v_1\tilde{v}_1/\sigma^2)=0.
\label{eext}
\end{equation}
It follows from Eq.~\ref{eext} that the distribution (Eq.~\ref{eb}) has
a local minimum at $v_1=0$ if $\tilde{v}_1/\sigma > \sqrt{2}$. Thus the
observation of such a minimum in the data is a sufficient condition for
the appearance of flow. But such a minimum is not required for the
extraction of $\tilde{v}_1$ and $\sigma$ by fitting Eq.~(\ref{eb}) to
data.

It will be shown in Section~\ref{smpt} that in the case of the analysis of
azimuthal $p_t$ distributions $\tilde{v}_1$
equals $N\langle p_x \rangle$, where $\langle p_x \rangle$ is
the mean transverse momentum gained by a particle due
to transverse flow ($x$-$z$ plane is the reaction plane here).
In this case
\begin{equation}
\frac{\tilde{v}_1}{\sigma}=
\frac{N \langle p_x \rangle}{\sqrt{N\langle p_t^2 \rangle /2}}=
\frac{\langle p_x \rangle \sqrt{2N}}{\sqrt{\langle p_t^2 \rangle}}.
\label{ervs}
\end{equation}
We see from Eq.(\ref{ervs}) that, one has to slice the rapidity windows
experimentally to accept reasonably large multiplicity ($N$) in each
window to make sure that $\tilde{v}_1/\sigma$ is not too small, and
an available statistics allows to perform a reliable fit. The
``relatively narrow'' rapidity windows discussed in the text
are precisely chosen this way.

Applying the same consideration to the distribution of the coefficients
corresponding to higher harmonics, one can see that they should have the
same behavior (Eqs.~$\ref{evar}-\ref{eb}$) but with their own parameters
$\tilde{v_i}$.
Note, that because of $\langle \cos^2(\phi) \rangle =\langle
\cos^2(n\phi) \rangle $ one has the same values of $\sigma$ for all
harmonics.

The functional form (\ref{eb}), which allows to have a local minimum of
the distribution at zero, was proposed almost a decade ago for the
distribution of flow polar angles extracted from the three dimensional
sphericity matrix~{\cite{lcu}}. For the two dimensional transverse
sphericity matrix a comprehensive study was done in Ref.~\cite{lo1,lo2}.
Here we want to emphasize that the same functional behavior with the
same parameter responsible for finite multiplicity fluctuations is valid
for the distribution of Fourier coefficients of any harmonic order.
Therefore one would expect more stable and sensible results fitting
simultaneously all $dw/v_ndv_n$ distributions with free parameters
$\tilde{v}_n$ ($n=1,2,3,4,...$) and the same parameter $\sigma$.

In order to be able to compare the magnitudes of flow in different
rapidity windows or centrality bins, or to compare different
experiments, it is useful to consider the values of $\tilde{v}_n$
normalized to $\tilde{v}_0=\langle x_0 \rangle=N\langle p_t \rangle$,
that is $\tilde{v}_n /\tilde{v}_0$.
The ratio $\tilde{v}_1 / \tilde{v}_0$ has a
clear physical meaning as the ratio of transverse momentum gained by a
particle due to transverse flow to the mean transverse momentum ($x$-$z$
plane being the reaction plane):
\begin{equation}
\frac{\tilde{v}_1  }{\tilde{v}_0}
=\frac{\langle p_x \rangle}{\langle p_t \rangle}.
\end{equation}

\section{Correlations} \label{sec4}

The parameter $\tilde{v}_n$ and $\sigma$ can be extracted from
experimental data by fitting the distributions $dw/v_n dv_n$ by the
functional form of Eq.~\ref{eb}. The non-zero $\tilde{v}_n$ reflects the
existence and magnitude of flow of ``$n$-th type''.  It is interesting
to study whether the detected flow in one rapidity window is correlated
with flow in a different rapidity window, or whether the flow of one
kind of particles is correlated with the flow of another particle type.
The study of some of such kind of correlations was proposed as proof of
an existence of flow in Ref.~\cite{ldo} (where the authors discuss in
our notation the correlations of $\psi_1$ from different rapidity
windows) and in Ref.~\cite{lo2} (correlations of $\psi_2$).  In our more
general approach along with these correlations we can study very
interesting correlations between flow of different types defining the
overall flow picture. In general, we can study the correlations
between any rapidity windows and any harmonics, $(v_m',m\psi_m')$ and
$(v_n,n\psi_n)$. Denoting the relative angle by $\phi=\psi_m'-\psi_n$
and averaging over the reaction plane angle we have the distribution:
\begin{eqnarray}
&&\frac{d^3 w}{dv_m'^2\,dv_n^2 \,d\phi} =
\int \frac{d^4 w}{dv_m'^2\,dv_n^2\,d\psi_m'\,d\psi_n}
d\psi_m'\,d\psi_n\,\delta (\phi-\psi_m'+\psi_n)=
\label{ec} \\
&& \, =
\frac{1}{4( 2 \pi\sigma^2)^2} \exp (-\frac{v_m'^2+\tilde{v}_m'^2}{2\sigma^2}
-\frac{v_n^2+\tilde{v}_n^2}{2\sigma^2})
\int_0^{2\pi} d\psi \, \exp (+\frac{v_m'\tilde{v}_m'}{\sigma^2}\cos(\psi)
+\frac{v_n\tilde{v}_n}{\sigma^2} \cos(\phi+\psi)), \nonumber
\end{eqnarray}
The observation of
correlation expressed in Eq.~\ref{ec} could be very helpful for
understanding the nature of flow and get a global picture of three
dimensional event shapes. The correlations between flow of different
types in the same rapidity window can be biased by auto-correlations due
to finite multiplicity.  To avoid the problem one can study the
correlations using as a reference a flow angle defined in another
rapidity window. The other way to suppress these pseudo-correlations could
be to calculate higher harmonic term after removal of lower harmonic
terms event-by-event, e.g., to perform a shift of the center of particle
distribution to the point
$(v_1,\psi_1)$ and then calculate $(v_2,2\psi_2)$.

The correlation between different rapidity windows can be numerically
calculated using the fit parameters to the distributions~(\ref{eb}). In
this case a comparison with experimental data will be a check of the
consistency of the method.  Note that in spite of rather complicated
form (\ref{ec}), the real calculation can be easily done numerically.
Some analytical calculations for specific parameters can be done using the
technique presented in the Appendix to Ref.~\cite{lo2}.

\section{Relation to other methods} \label{sec5}

\subsection{Directivity and sphericity}

The vector ${\bf Q}$, which is often referred as directivity, was
originally introduced in the flow analysis by Danielewicz and Odyniec
{\cite{ldo}}:
\begin{equation}
{\bf Q}=\sum_{\nu=1}^N \omega_{\nu} {\bf p}_{t;\nu},
\end{equation}
where $\nu$ is particle index and $\omega_{\nu}$ is a weight.  It
coincides with the vector ${\bf v}_1\equiv (v_1,\psi_1)$
in the Fourier analysis by taking $\omega_{\nu}=1$.  One of the
results of Ref.~\cite{ldo} concerning the width of the distribution in
${\bf Q}$ is directly related to the modification of the original
Gaussian distribution due to flow, which we discussed above.

The relation of the second harmonic coefficients to the usual
parameterization of two dimensional sphericity tensor is also
straightforward.  Two dimensional transverse sphericity tensor, which is
relevant to our study, has three independent parameters, and usually is
written in the form:
\begin{equation}
S_{ij} \equiv \sum_{\nu =1}^N \omega_{\nu} p_{i;\nu}
                p_{j;\nu}=\frac{S}{2}
    \left( \begin{array}{cc}
    1+\epsilon \cos 2\theta & \epsilon \sin 2 \theta \\
                        \epsilon \sin 2 \theta & 1-\epsilon \cos 2\theta
    \end{array}  \right),
\end{equation}
where $i,j=x,y$. If we chose $\omega_{\nu}=r_{\nu}/(p_{t;\nu})^2$
the sphericity tensor becomes
\begin{equation}
S_{ij}=\sum_{\nu =1}^N r_{\nu} \cos \phi_{\nu ,i} \cos \phi_{\nu ,j},
\end{equation}
where $\phi_{\nu ,i}$ is an angle between the momentum of $\nu$-th
particle and $i$-th axis.  In this case $S={\rm tr} S_{ij}=\sum_{\nu}
r_{\nu} =v_0$.  By computing the eigenvalues one can derive that the
eccentricity $\epsilon = v_2 /v_0$ and thus is directly connected to the
magnitude of the second harmonic coefficient in Fourier analysis.

\subsection{Reaction plane determination}  \label{srp}

Experimentally, to study the flow in the reaction plane (so called
side-splash flow), traditionally a directivity (an angle $\psi_1$) is
calculated from the whole available acceptance to define the reaction
plane. To define the accuracy of the procedure one should consider the
distribution in an angle $\psi$ which is defined as $\psi=\psi_r-\psi_1$:
\begin{eqnarray}
\frac{dw}{d\psi}
&=&
\int \frac{vdv}{2\pi \sigma^2}
\exp (-\frac{\tilde{v}_1^2+v_1^2-2\tilde{v}_1 v_1 \cos \psi}{2\sigma^2})=
\nonumber
\\
&=&\frac{1}{2\pi}[
\exp (-\frac{\tilde{v}_1^2}{2\sigma^2})
+
\frac{\tilde{v}_1 \cos \psi \sqrt{\pi}}{\sigma \sqrt{2}}
\exp (-\frac{\tilde{v}_1^2 \sin^2\psi}{2\sigma^2})
(1+{\rm erf}(\frac{\tilde{v}_1 \cos \psi}{\sigma \sqrt{2}}))].
\label{edpsi}
\end{eqnarray}
As long as the parameters of flow are determined by fitting the
experimental distribution to the form (\ref{eb}), Eq.(\ref{edpsi}) can
be used to answer the question of how close the angle $\psi_1$ is to the
true reaction plane angle $\psi_r$. It can be generalized to $n$-th type
flow angle dispersion by replacing $v_1$ with $v_n$, and $\psi_1$ with
$n\psi_n$.

\subsection{Mean $p_x$}  \label{smpt}

Any transverse flow would influence the $p_t$ distribution of observed
particles.  Theoretical calculations~{\cite{lam}} usually present a mean
projection of the particle transverse momentum into the reaction plane
$\langle p_x \rangle$ as a function of rapidity
(it is assumed here that $x$-$z$ plane is the reaction plane).
The projection can be non-zero due to
transverse flow.  Generally, as discussed above, the directivity is used
to define the reaction plane, and then in different rapidity bins
transverse momenta are projected into this plane.  (One should be
careful here to avoid autocorrelation.  For details see
Ref.~\cite{ldo}.)  Below we use the notation $p_{x'}$ for such a
projection instead of $p_x$ to note that it is not the projection into
true reaction plane.  Due to the fluctuations in $\psi_1$ around the
true reaction plane angle  $\langle
p_{x'}\rangle$ does not equal $\langle p_x\rangle$.
In our approach, we get our
fitting parameter $\tilde{v}_1$ without determining of the reaction
plane on an event-by-event basis.  Taking into account that
$\displaystyle {\bf v}_1=\sum_{\nu}{\bf p}_{t;\nu}$, one can see that
$\tilde{v}_1$ is exactly the true $\langle p_x\rangle$, multiplied by
the total number of particles in the rapidity region.
\begin{equation}
\langle p_x\rangle \,N =
 	\frac{1}{2\pi \sigma^2}
%\int_0^{2\pi} \frac{d\psi_r}{2\pi}
        \int d{\bf v}_1
      v_1 \cos (\widehat{{\bf v}_1,\tilde{{\bf v}}_1})
   \exp (-\frac{({{\bf v}_1}-\tilde{{\bf v}}_1)^2}{ 2 \sigma^2})=
    \tilde{v}_1.
\end{equation}
The relation between $\langle p_x \rangle$ and $\langle p_{x'} \rangle$ is
very simple~{\cite{ldo}}
$\langle p_x \rangle
= \langle p_{x'} \rangle \langle\cos (\psi_{xx'})\rangle$, where
$\psi_{xx'}$ is the difference between true and defined reaction plane
angles $\psi_{xx'}=\psi_r -\psi_1$. In our approach
\begin{eqnarray}
\langle \cos (\psi_{xx'}) \rangle &=&
 	\frac{1}{2\pi \sigma^2}
% \int_0^{2\pi}\frac{d\psi_r}{2\pi}
        \int d{\bf v}_1
	\cos (\widehat{{\bf v}_1,\tilde{{\bf v}}_1})
      \exp (-\frac{({\bf v}_1-\tilde{{\bf v}}_1)^2}
	{ 2 \sigma^2})= \nonumber \\
 &=&  \frac{1}{\sqrt{2\pi}}
      \frac{\tilde{v}_1}{\sigma}
     \exp (-\frac{\tilde{v}_1^2}{2\sigma^2})
\, _1F_1(3/2;2;\frac{\tilde{v}_1^2}{2\sigma^2}),
\end{eqnarray}
where $_1F_1(\alpha ;\beta ;\gamma)$ is a confluent hypergeometric
function.
The behavior of $\langle \cos (\psi_{xx'}) \rangle$ as a function of
$\tilde{v}_1/\sigma$ is shown in Fig.\ref{fmeancos}.

\subsection{Azimuthal two-particle correlations}

An azimuthal two-particle correlation function
$C(\psi)=P_{cor}(\psi)/P_{uncor}(\psi)$, where $\psi$ is the angle
between two particles, was proposed in Refs.~\cite{lcf,lcf2} to study
transverse flow as a method which is free from uncertainty in the
reaction plane determination.
Here $P_{cor}(\psi)$ is observed pair distribution and
$P_{uncor}(\psi)$ is the distribution generated by ``event mixing''.
The formalism used above for Fourier analysis permits
us to estimate the azimuthal two-particle correlations in the same way
as it was done for the distributions of Fourier coefficients.  The
origin of azimuthal correlations lies in the overall shift of the
particle distribution in ${\bf p}_t$-space due to flow.  The parameter
which controls the magnitude of the correlations is $\langle p_x
\rangle/\sqrt{\langle p_t^2 \rangle}$.
The correlations increase with increasing
values of this parameter.  Note that the corresponding parameter (which
defines the magnitude of the effect) in our approach is
$\tilde{v}_1/\sigma
=N\langle p_x \rangle(y)/\sqrt{N\langle p_t^2 \rangle/2}$, which is
$\sqrt{N}$ times larger.  The difference, as always, originates from the
fact that in the study of collective phenomena the fluctuations in the
collective variables are less than the fluctuations in variables describing
individual particles.

If one assumes a Gaussian form for the one-particle $p_t$ distribution,
the expression for $C(\psi)$ coincides with the expression for the
correlation between two flow angles defined in different regions, after
a substitution of corresponding parameters.  The statement made above
about finite multiplicity fluctuations becomes explicit in this case.

\section{Conclusion} \label{sec6}

To summarize we propose to study transverse anisotropic collective flow
by analysis of azimuthal distributions in different rapidity regions.
We show that Fourier expansion of the azimuthal distributions is an
appropriate tool for such an analysis.  It is shown that the
distributions in $v_n=\sqrt{x_n^2+y_n^2}$, where $x_n$ and $y_n$ are
$n-$th harmonic Fourier coefficients, are sensitive to transverse flow.
The classification of flow of different types leading to a deformation
of azimuthal distribution corresponding to different Fourier harmonics
appears naturally in this method. The important feature of the approach
is that it is free from uncertainties in a determination of the reaction
plane on an event-by-event basis. The dispersion of the flow angles is
derived through the discussion of the finite multiplicity fluctuations.
Correlations in the flow angles determined for flow of different types
are proposed as a technique for flow analysis. It is shown that many
existing methods are within our framework, either the first or the
second harmonics. But our approach provide additional information, from
which a three dimensional event shape can be established.

\acknowledgments

The authors have gained a lot from fruitful discussions with
P.~Braun-Munzinger, W.~Cleland, J.~Stachel and J.P.~Wessels. One of the
authors (YZ) would like to thank N.~Herrmann for some initial
discussions.  Financial support was provided by the NSF, and the US
Department of Energy under Contract No. DEFG02-87ER40363.

\clearpage
\newpage
\section*{Figure captions}

\begin{enumerate}

\item     \label{ffl}
Function $r (\phi )$ (solid line). Approximation by ellipse (dashed line).

\item   \label{feb}
The distribution $dw/xdx$ for different
values of $\tilde{x}$. The curves marked by 1,2,3, and 4 correspond to
$\tilde{x}=0.5, \,1.0, \,1.5, \,2.0$, respectively.

\item \label{fmeancos}
$\langle \cos (\phi_{xx'}) \rangle$ as a function of $\tilde{v}_1/\sigma_s$
\end{enumerate}

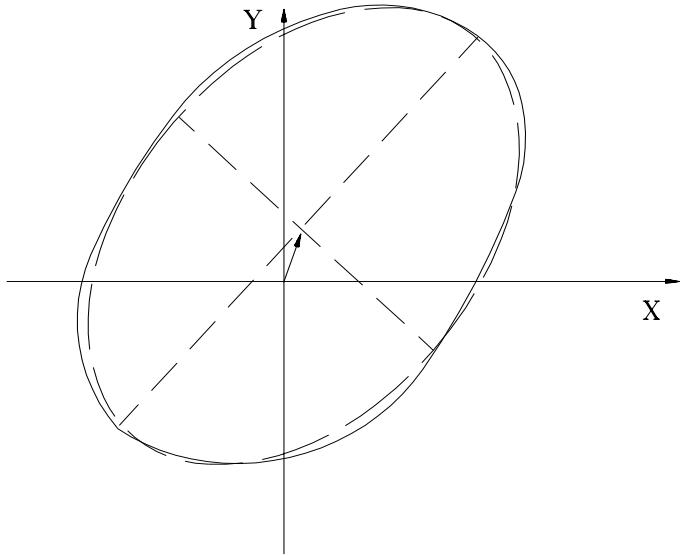
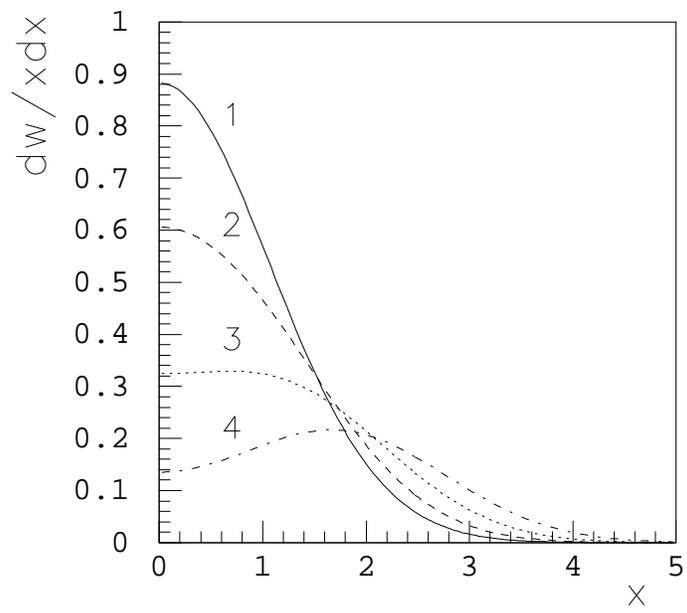
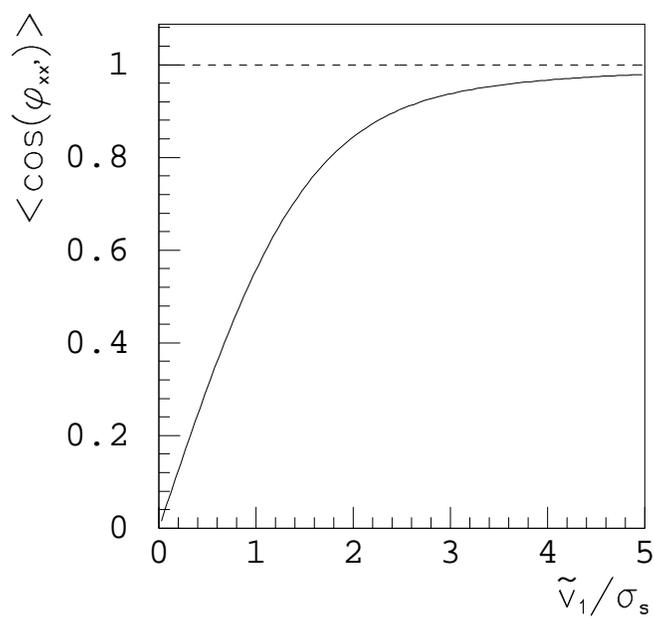
\end{document}